\documentclass[letterpaper]{article} 
\usepackage{aaai2027}  
\usepackage[hyphens]{url}  
\usepackage{graphicx} 
\usepackage{natbib}  
\usepackage{caption} 
\usepackage{algorithm}
\usepackage{algorithmic}
\usepackage{tabularx} 
\usepackage{newfloat}
\usepackage{listings}
\DeclareCaptionStyle{ruled}{labelfont=normalfont,labelsep=colon,strut=off} 
\floatstyle{ruled}
\newfloat{listing}{tb}{lst}{}
\floatname{listing}{Listing}

\usepackage{booktabs}

\title{Found in Translation: Five Entry Points to Translate AI Literacy Frameworks\\ in the Design of Educational Materials}
\author{
    Luis Morales-Navarro,\textsuperscript{\rm 1}
    Karl-Emil K. Bilstrup,\textsuperscript{\rm 1, 2}
    Mille Skovhus Lunding,\textsuperscript{\rm 1}
    Luke Connelly,\textsuperscript{\rm 1}
    Ane V.	Jensen,\textsuperscript{\rm 1}
    Maja Dybboe,\textsuperscript{\rm 1}
    Yasmin B. Kafai,\textsuperscript{\rm 3}
    Marianne Graves Petersen,\textsuperscript{\rm 1}
    Ole Sejer Iversen\textsuperscript{\rm 1}
}
\affiliations{
    \textsuperscript{\rm 1}Aarhus University, Aarhus, Denmark\\
    \textsuperscript{\rm 2}University of Copenhagen, Copenhagen, Denmark\\
    \textsuperscript{\rm 3}University of Pennsylvania, Philadelphia, USA

    luismn@cc.au.dk;
    keb@di.ku.dk;
    milledsk@cs.au.dk;
    luke@cs.au.dk;
    av@cc.au.dk;
    dybboe@cs.au.dk;
    kafai@upenn.edu;
    mgraves@cs.au.dk;
    oiversen@cc.au.dk
}

\begin{document}

\maketitle

\begin{abstract}
While there is a plethora of AI literacy and informatics frameworks, in practice it is difficult for designers of  educational materials and educators to decide where to start and how to use these in learning activities. At the same time, recent reviews show that despite the fact that most frameworks emphasize the socio-technical nature of AI, in practice interventions tend to center on preparing learners to use AI. To address these issues, we ask: How can multiple stakeholders negotiate and translate AI literacy framework ideas into common language that can be used to design introductory learning materials for primary school children, teachers and parents? We introduce five entry points into AI literacy that were developed from a computational empowerment perspective through a participatory design process involving primary teacher educators, curricular designers, media producers, and researchers. We present three vignettes that illustrate how the entry points were generative in the design of learning activities, learning tools and media content for school and home. This paper contributes an account of the participatory development of five entry points that serve as generative and negotiable resources for designing educational materials.

\end{abstract}

\section{Introduction}

\begin{figure*}
    \centering
    \includegraphics[width=1\linewidth]{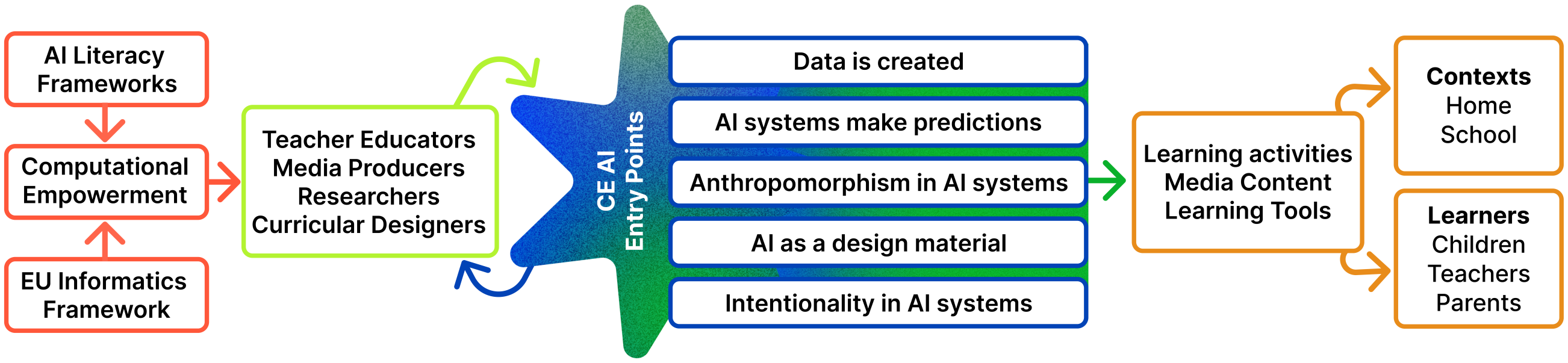}
    \caption{Participatory design process diagram}
    \label{fig:placeholder}
\end{figure*}

As artificial intelligence continues being integrated into everyday life and society at large, increasing attention has been given to young people's AI literacy in order for them to ``to navigate the presence of AI in daily life and to develop
skills to use it ethically and creatively'' \cite{OECD2026EmpoweringLearners}. Within the AAAI community, efforts guided by the five big ideas \cite{touretzky2019envisioning} have highlighted the need to support everyone from children to older adults in 
developing AI literacy \cite{xu2026essentials, kasinidou2026educating}. However, many AI literacy efforts, particularly in relation to generative AI, fundamentally lack a critical and reflective stance towards technology \cite{10.1145/3702652.3744217}. Furthermore, many efforts center on supporting people to learn how to \textit{use} AI~\cite{kalantzis2025literacy}, lacking an agenda of empowering children to understand and participate in AI and make informed decisions about their relationships to technology. 

We build on the ideas of computational empowerment (CE)~\cite{10.1145/3210586.3210592} to argue that children should go beyond learning how to use AI. Computational empowerment was born out of the Scandinavian participatory design tradition as a response to computational thinking efforts that focused on coding. Centered on children's lived experiences, computational empowerment is ``the process in which children, as individuals and groups, develop the skills, insights and reflexivity needed to understand digital technology and its effect on their lives and society at large, and their capacity to engage critically, curiously and constructively with the construction and deconstruction of technology''~\cite{10.1145/3210586.3210592}. Approaching AI literacy from a CE perspective shifts the focus from learning how to use AI and conceptual understanding of AI to providing children with opportunities to participate in the development and evaluation of AI systems. 

While AI literacy frameworks and broader computing education frameworks (e.g., the EU Informatics for All framework \cite{InformaticsForAll2022Framework}) are helpful in guiding the content of AI literacy interventions, in practice it is difficult for the designers of educational materials and educators to decide where to start \cite{yang2026systematic}. As \citet{casal2023ai} show, enacting frameworks in learning activities remains a challenge, as these are rarely adapted to the specific contexts of school. Designers of learning materials indeed can be overwhelmed in choosing among frameworks. To address this issue, in this paper, we ask: \textbf{How can multiple stakeholders negotiate and translate AI literacy framework ideas into common language that can be used to design introductory learning materials for primary school children, teachers and parents?} We introduce five \textit{entry points} into AI literacy that were developed through a participatory design process involving diverse stakeholders, including researchers, teacher educators, curricular designers, and media producers. We define entry points as conceptually grounded starting points for the design of AI literacy materials. This paper shows how these co-designed entry points supported different actors to maintain a shared, generative foundation while designing diverse AI literacy materials.

To illustrate how the entry points can guide AI literacy efforts, we present three vignettes that demonstrate how we have used them in the processes of designing learning materials, writing screenplays for a children's TV show, and creating learning activities for parents and children. Based on these vignettes and our broader experiences from the project, we discuss how the entry points facilitated translating framework ideas and using these in the design of learning materials. We contribute entry points that support bridging the gap between AI literacy and informatics frameworks and the design of educational materials and, in addition, serve as shared  reference points in collaborative design processes. We stress entry points as being generative statements that are plastic and negotiated by stakeholders. Moreover, the five specific entry points offer a starting point for AI literacy from a computational empowerment perspective.

\section{AI Literacy and Computational Empowerment}
AI literacy continues to gain traction and attention from educators and policymakers with little agreement on what it means \cite{10.1145/3626252.3630937}. A recent definition from the \citet{OECD2026EmpoweringLearners}, wherein this research is conducted, states:
\begin{quote}
``AI literacy represents the technical
knowledge, durable skills and
future-ready attitudes required to
thrive in a world influenced by AI. It
enables learners to engage, create
with, manage and shape AI, while
critically evaluating its benefits, risks
and ethical implications.''
\end{quote}

Such a definition puts value on both technical and socio-ethical aspects of AI. This is not a surprise since most AI literacy frameworks, from the five big ideas \cite{touretzky2019envisioning}, to the United Kingdom's SEAME framework \cite{waite2023ethics}, to \citeauthor{10.1145/3313831.3376727}'s \citeyearpar{10.1145/3313831.3376727} competencies and the UNESCO framework \cite{unesco2024aicompetency}, emphasize socio-ethical aspects of AI. However, a recent review of the literature shows that AI literacy interventions centered around generative AI lack a critical and reflective stance towards technology \cite{10.1145/3702652.3744217}. Oftentimes, such interventions also lack a technical stance towards technology, centering on preparing young people to \textit{use} AI tools rather than to understand how AI systems work \cite{kalantzis2025literacy}. Furthermore, while literacy frameworks are useful in guiding AI education efforts, there is little evidence of how these are put into action. One example of an approach that does focus on practice is \citet{10.1145/3769994.3770026} who articulate a pedagogical approach to introduce AI literacy ideas in K-12. 

Whereas policymakers around the world continue to debate how to prepare young people for a present and future with AI by introducing AI literacy and to discuss which frameworks to adopt, enacting frameworks can be a challenge. Researchers have documented that such enactment can be difficult due to knowledge gaps between framework authors and designers of learning materials and educators \cite{yang2026systematic} and lack of clear guidance that accounts for local contexts \cite{casal2023ai, mutawa2025unesco}. AI literacy interventions for primary school children also face challenges as frameworks tend to emphasize concepts that are applicable for older children, centering on middle school and high school \cite{yeter2024global}. Furthermore, the wide range of concepts and applications can be overwhelming for teachers and designers to grasp and make decisions about where to start. 

Beyond the challenge of where to start, we argue that we should critically consider what the goals of such literacy efforts are and how distinct interventions support such goals. AI literacy in the broadest sense includes important skills and competences needed for a contemporary labor market, including knowing how to use and operate AI systems. However, learning job skills and how to use AI cannot truly prepare learners to fully participate and make decisions about the role of AI in their lives. A computational empowerment approach 
centers on how children develop understanding of the technologies they encounter in their everyday lives, transformative agency to participate in a socio-technical world, and critical hope to imagine better futures. CE is fostered through construction and deconstruction activities in which learners are protagonists or the ``main agents'' in designing and evaluating systems \citep{iversen2017child}.


We argue that CE has a lot to offer in guiding AI literacy interventions. First, CE centers on children's lived experiences, making connections to the everyday interactions that children have with AI. Second, CE addresses both technical and socio-ethical aspects of computing through the construction and deconstruction of technology. Third, CE is expansive, going beyond learning in the classroom. In this paper, we describe an AI literacy project that aims to support children, teachers and parents in learning about AI at home and in school through learning activities, learning tools and media content. In particular, we center on how different stakeholders negotiated and translated ideas from frameworks into entry points that could facilitate the use of framework concepts in the design of introductory learning materials for primary school children, teachers and parents.
\section{Participatory Design Process}
This paper describes an ongoing participatory design (PD) process within the Ultra:Bytes project, from which the five entry points have emerged. Ultra:Bytes is a multi-partner project that aims to strengthen AI literacy among K-6 children in Denmark. It brings together a national public-service media company; a teacher education college supporting educators, schools, and municipalities with teacher education; a private foundation supporting STEM education research and young people's motivation in STEM; and two interdisciplinary research groups, one based in Northern Europe and one in the Northeastern United States, with expertise in AI, participatory design, and learning sciences.

The decision to organize this work through PD follows the project's computational empowerment orientation. CE is rooted in PD because it connects understanding technology with the capacity to question and shape it (Iversen, Smith, and Dindler 2018). In this project, this principle also applies to the professionals who design the media content, learning technologies, and learning activities through which children and their teachers and parents encounter AI. Following \citet{958052.958061}, participation was not treated as simply introducing these stakeholders to new AI concepts. The aim in the project is to enable them to understand AI and AI literacy well enough to question our shared AI conceptions and decide how AI literacy frameworks should inform their own work. We therefore considered stakeholder ownership of the \textit{entry points} necessary for them to have a meaningful frame for developing AI literacy learning materials throughout the project.

Before the first partner workshop, the European research team compared the Five Big Ideas in AI \cite{touretzky2019envisioning}, the AI literacy competencies proposed by \citet{10.1145/3313831.3376727}, and the 11 core topic areas of the Informatics Reference Framework for School \cite{InformaticsForAll2022Framework}. This comparison identified conceptual overlaps and produced a set of provisional topic areas for AI literacy in K-6 education. With a CE lens in our work, we prioritized topics that connected technical and socio-ethical perspectives and could be related to children's everyday experiences with AI. The European and US-based research teams then consolidated and refined these topics through a series of online workshops.

At the initial partner PD workshop, the researchers presented the provisional topic areas as starting points for discussion, not as a fixed specification of what the project should produce. Media producers, teacher educators, and learning-technology designers examined the topics from the perspective of their own professional practices. They considered how each topic might be expressed through narratives and streaming content, learning tools, or facilitated learning activities for K-6 children. The discussion did more than translate predefined AI concepts into different formats: the stakeholders' questions, priorities, and practical experience also changed how the topics were worded, emphasized, and related to one another. The first working formulations of the entry points emerged through this exchange between the researchers' provisional topics and the stakeholders' expertise. These formulations were then used in subsequent workshops and development activities. Table 1 shows how the resulting entry points developed from this process can be traced back to the three frameworks.


As partners used the entry points to develop media content, learning technologies, and educational activities, ambiguities and different interpretations prompted further negotiation within the research teams and between researchers and stakeholders. Through this process, the entry points took their current form and came to function as flexible and common objects \cite{star1989structure}: recognizable across stakeholders, yet adaptable to each professional practice. Because partners could adjust rather than simply receive the entry points, it supported coordination and shared ownership. Consequently, the entry points became both products and means of ongoing negotiation in the project. This became apparent in a mapping exercise in which different stakeholders mapped how they adapted and adopted the entry points to use them in the design of learning materials. 

In the following sections, we first present the five entry points in their current form and then show how they have been used in ongoing design work. We do not argue that these five entry points alone are sufficient to foster AI literacy or that they constitute a universal framework. Our contribution is an account of how these entry points, negotiated with professional expertise and kept open to revision, can support the collaborative development of materials for K-6 children with an emphasis on computational empowerment. The plasticity of the entry points is central to this role. It enables designers, media professionals, researchers and teacher educators to have a shared generative starting point to foster AI literacy. 
\section{Five Entry Points}
Here we describe these five entry points, provide examples from existing research on CE, and highlight their relationship to existing frameworks.



\renewcommand{\arraystretch}{1.3}
\begin{table*}[htbp]
\centering
\renewcommand{\arraystretch}{1.3} 

\begin{tabularx}{\textwidth}{>{\raggedright\arraybackslash}p{3.2cm} 
                             >{\raggedright\arraybackslash}X 
                             >{\raggedright\arraybackslash}X 
                             >{\raggedright\arraybackslash}X}
\toprule
\textbf{Entry point} & 
\textbf{Five Big Ideas} \newline\cite{touretzky2019envisioning} & 
\textbf{AI Competences} \newline\cite{10.1145/3313831.3376727} & 
\textbf{Informatics Reference Framework Topics} \cite{InformaticsForAll2022Framework} \\ 
\midrule

Data created & 
3. Learning & 
12. Learning from Data\newline 13. Critically Interpreting Data & 
Data and information \\

AI systems make predictions & 
3. Learning & 
9. ML Steps\newline 12. Learning from Data & 
Modeling and simulation \\

Anthropomorphism in AI systems & 
4. Interaction\newline 5. Impact & 
5. AI's Strengths and Weaknesses\newline 16. Ethics & 
Human-machine interaction \\

AI as a design material & 
& 
10. Human Role in AI & 
Digital creativity \\

Intentionality in AI systems & 
& 
10. Human Role in AI & 
Responsibility and Empowerment \\

\bottomrule
\end{tabularx}
\caption{Entry Points and Related Frameworks}
\label{tab:entry}
\end{table*}

\subsection{Data is created}
Data are representations that are created to describe phenomena in the world. This entry point centers on data as an artifact that is actively produced by humans.
Children use AI systems when playing games, learning, and listening to music; however, their interactions with data are often obscured, and data is often framed as something that is passively collected instead of actively created.
Within CE, this entry point has been previously explored, for example, by having learners produce and analyze personally meaningful music data and cocreate playlists while critically reflecting on data practices \cite{10.1145/3772318.3791296}. Other work has involved learners in examining the datasets of faulty AI systems \cite{11574161}. As \citet{hardy2020data} argued, engaging learners in producing data for personally relevant purposes can support them to develop their agency. This entry point relates to Big Idea 3, Competencies 12 and 13, and the Data and Information core topic of the EU's framework.

\subsection{AI systems make predictions}
This entry point refers to how AI systems make predictions based on patterns identified in the data on which they are trained. 
While children interact with AI systems daily, oftentimes outputs are perceived as deterministic and trustworthy. Yet it is important for children to understand that outputs are predictions made based on prior data.
Work in CE has explored this entry point by having learners create datasets and train models \cite{10.1145/3547522.3547689}, exploring how Markov chains make predictions \cite{10.1145/3773077.3806141}, and analyzing the outputs of generative AI systems \cite{10.1145/3757620}. In these activities, learners engage with the non-deterministic nature of AI and develop some understanding about how data quality shapes the outputs of AI systems. This entry point relates to Big Idea 3, Competencies 9 and 12, and the Modeling and Simulation core topic.

\subsection{Anthropomorphism in AI systems}
Anthropomorphism refers to the tendency to attribute human characteristics, such as empathy, intention, and understanding, to AI. This occurs even though AI systems compute predictions are based on data, and cannot think or understand in the same way humans do. 
Yet, recent studies show that people increasingly perceive AI systems as human-like~\cite{cheng2026metaphors}. Chatbot interfaces and consumer products that target children (e.g., AI toys such as Curio) replicate deceptive human-like characteristics. 
While anthropomorphism has not been widely addressed in CE literature, research shows that it is pervasive in the design of interfaces for children \cite{10.1145/3773077.3806126} and it has been argued to contribute to misunderstandings and obstruct students' learning about AI if not addressed in learning activities~\cite{Whyte2024_secondary_AI_conception}. This entry point relates to Competency 10 and the human-machine interaction topic of the EU's framework.

\subsection{AI as a design material}
AI as a design material refers to how AI systems are designed by humans.
While children engage with AI systems, these may be perceived as autonomous, intelligent, and/or magical. 
We argue that learners can participate in designing AI systems by creating and modifying them, defining what models should make predictions about, designing datasets, and choosing parameters, as well as designing the interfaces that people use to interact with AI. Within CE, research shows that framing AI as a design material can support learners to understand the possibilities and limitations of AI \cite{10.1145/3547522.3547689, kahila2026engaging}. Several studies show that when learners participate in designing AI systems, their perceptions shift from seeing these as autonomous to understanding them as human-designed artifacts \cite{10.1145/3459990.3460712, 10.1145/3773077.3806107}. This entry point relates to Competency 10 and the Digital Creativity topic of the EU's framework.

\subsection{Intentionality in AI systems}
Intentionality concerns the choices, values, and purposes that humans embed in AI technologies through design. This entry point highlights that AI is therefore not neutral; rather, it reflects the intentions and priorities from which these technologies are created. 
Children encounter technology at home and in school, and adults often have opinions about it. However, the intentions and values behind how such technologies were designed are not always clear for children and the adults around them.
CE research has involved learners in reflecting on the values and ethical stances embedded within AI systems \cite{10.1145/3357236.3395558}, this work has also highlighted how considering values and intentionality should be part of the process of designing AI systems \cite{10.1145/3531146.3534634}. Other research has engaged learners in reflecting on the real-world consequences of such design decisions \cite{hilke2026breakable}. This entry point relates to Competency 10 and the Responsibility and Empowerment topic of the EU's framework.

\subsection{The relationship between entry points}

The five entry points share a human-centered socio-technical perspective on AI that highlights the role of humans as designers of AI systems and producers of data that make decisions and embed values and intentions into AI systems. This is not a coincidence since computational empowerment research has a commitment to making transparent how technology ``is constructed and the consequences that it has'' \cite{dindler2022computational}. The entry points are nonhierarchical in that any of them could be used as a starting point in the design of educational materials. They are also non-exclusive; educational materials can address one or more of them. These are starting points for design, so it is even possible to arrive at similar activities from different entry points.
\section{Vignettes}
We use narrative vignettes \cite{erickson2012qualitative} to portray how the entry points are enacted in the context of the project. The vignettes were iteratively developed in four steps: the authors identified focal episodes in the participatory design process, gathered transcripts and documents of such events, collaboratively discussed the entry points enacted, and wrote the vignettes.

\begin{figure}
    \centering
    \includegraphics[width=1\linewidth]{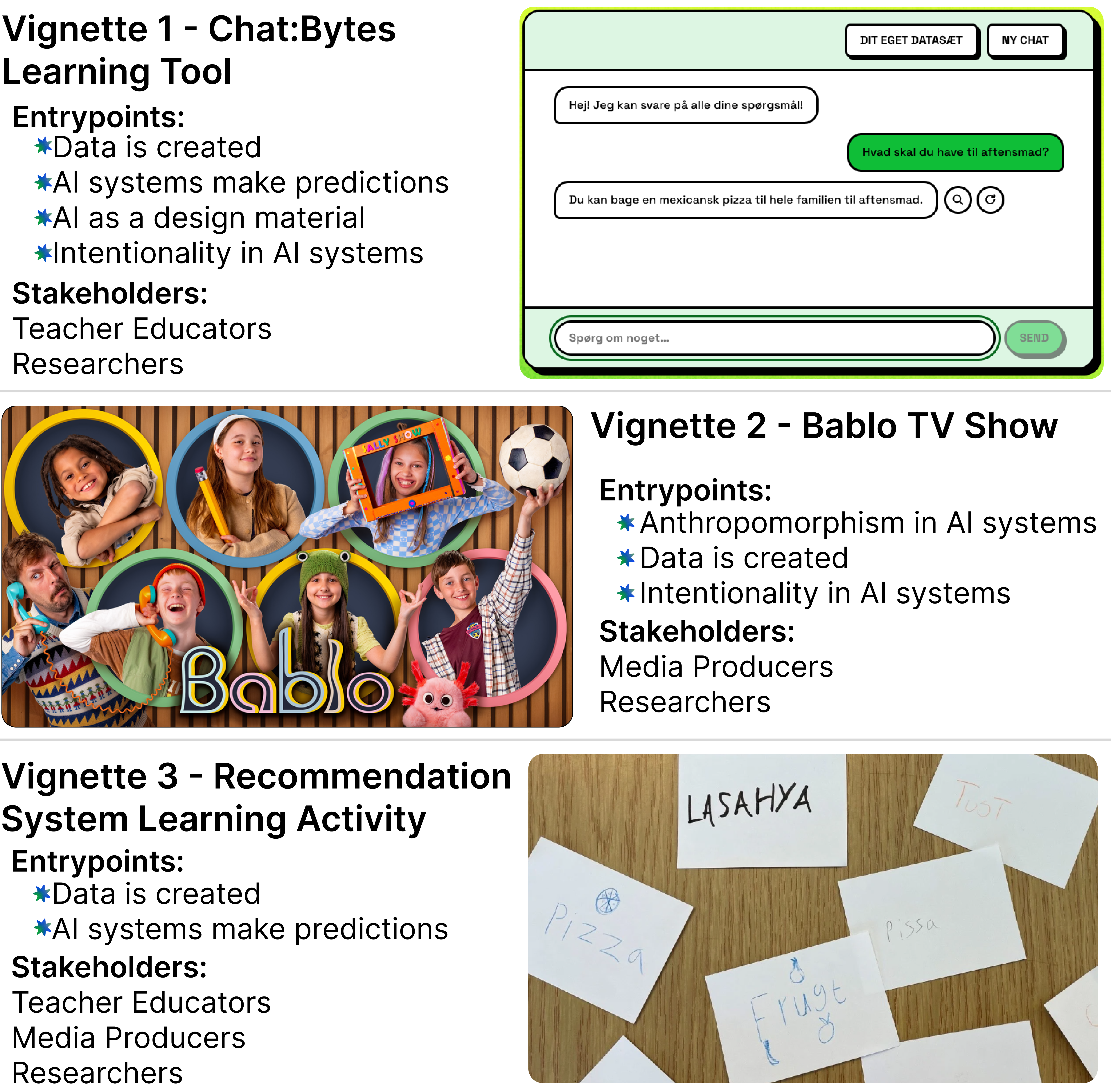}
    \caption{Examples of the entry points used by different stakeholders to create a learning tool, a tv show, and a learning activity}
    \label{fig:vignettes}
\end{figure}

\subsection{Vignette 1: Designing a Learning Tool for School}

As AI literacy has not been formally introduced into Denmark's K-6 curriculum, teacher educators have limited existing resources for introducing AI. In particular, a group of educators designing learning activities for children aged 10-12 wanted to create activities to foster understanding of how chatbots are designed and generate outputs. They wanted these activities to relate to the everyday experiences that children may have with commercial chatbots. The resources available were either too black-boxed or too complex for K-6 students. The educators wanted a new learning tool that could close this gap and provide some transparency into how chatbots work. In particular, they wanted a tool that could make transparent that \textbf{data is created} and that \textbf{AI systems make predictions}. Using those starting points, the challenge was addressed in a collaboration with researchers through the design of Chat:Bytes (see Figure \ref{fig:vignettes}). Chat:Bytes was developed by researchers but iteratively evaluated with teacher educators to assess how the entry points were manifested into the design.

From a technical perspective, Chat:Bytes combines a selection of classic NLP techniques (whitespace tokenisation, TF-IDF scoring, Markov Chains) and packages them into a chat interface that allows users to prompt and receive explainable responses. In deciding the features of the tool the educators and researchers agreed that Chat:Bytes should be designed to highlight how \textbf{data is created} and how it influences the answers that chatbots provide. Chat:Bytes was created to use only a small dataset of around 100 sentences. The interface made the dataset visible and explorable so that students could directly see how training data influences the answers of Chat:Bytes. The explorable dataset and data visualization of the predictions made by the chatbot increased transparency and made the entry point \textbf{AI systems make predictions} tangible. Furthermore, the content of the dataset represented controversial views, for example, that cats are better pets than dogs, to enable discussions about how the data created for an AI system can have implications in the outputs users encounter. 

As the tool developed, the design team wondered if this was an opportunity to support learners in engaging with \textbf{AI as a design material}. This led the team to brainstorm how the tool could support learners to modify the chatbot and embed their own \textbf{intentions}. To allow students to influence the design of Chat:Bytes, the interface was designed so that students could add new sentences to the training data and thereby influence the chatbot to answer differently about specific topics or add data about new topics to enable Chat:Bytes to provide  meaningful answers. By allowing students to add additional training data to Chat:Bytes, it also enabled them to embed their own \textbf{intentions}. For example, in a design meeting, Thomas
, a teacher educator, showed how he could change the intention of the chatbot by adding training data that included negative opinions about cats and positive opinions about dogs. In this vignette, we showed how working together teacher educators and researchers used the entry points as starting points to design a tool and learning activities around it.

\subsection{Vignette 2: Creating Media Content for the Home}

The Danish Broadcasting Corporation (DR), a partner in the project, decided to create 12 8-minute episodes of a TV show targeting children ages 7 to 11 called \textit{Bablo}. 
This is a fictional live-action TV show that happens in a library where children encounter problems and go on adventures. Bablo is not explicitly about AI, each episode has a hidden ``AI message'' such that it can be used in classrooms to discuss that message. 

A managing editor at the media company used the entry points, for instance \textbf{anthropomorphism in AI systems}, and \textbf{data is created}, to draft a set of messages about AI that were discussed and iterated together with researchers. The screenwriters then wrote initial drafts for possible storylines. These storylines were also reviewed by researchers who  provided feedback before the screenwriters wrote them into full-fledged screenplays for the episodes. Below we provide examples from two screenplays to demonstrate how the entry points supported the writing process.

In the first example episode, the idea was to provide children with a story about \textbf{anthropomorphism}. In this episode Malini is playing with Sally’s stuffed toy, LotteAxel. LotteAxel is described by the children as being upset by this. At the end of the episode, Sally admits that it may be her feelings, and not LotteAxel's, that are getting hurt. In the episode, Mailini gets very concerned. Sally then explains that ``It's just a game [...] LotteAxel doesn't actually have feelings, you know. She's just a stuffed animal.'' Sally is surprised by this, as Mailini always talked about the stuffed toy as if it were a human with feelings. The episode was written to convey that humans have a tendency to \textbf{anthropomorphize} non-humans by giving them human traits. It highlights how Sally attributes human feelings to the toy to serve as a conversation starter in the classroom about how humans anthropomorphize toys and, with the guidance of teachers, make connections to how humans give human-like characteristics to AI systems. 

In a second example episode, the producer, researchers and screenplay writers agreed to highlight how \textbf{data is created} and how, in creating datasets, designers embed their own \textbf{intentionality}. In this episode Bo and Sally are very bored when Sally’s bigger sister, Anna, brings an old gumball machine that is full of game suggestions written on post-its. However, Bo and Sally do not enjoy the games because the machine is filled up with Anna's favorite games. At the end of the episode, Bo and Sally realize they can add their own post-it notes to the gum machine. This show, without explicitly addressing AI, was designed to be used in the classroom to reflect on how \textbf{data is created}.

\subsection{Vignette 3: Designing a Recommendation System Learning Activity Bridging School and Home contexts}

At the same time as the broadcasting company screenwriters were writing the episodes described above, the educators started creating learning activities that could be used at school and at home. By using the \textbf{data is created}, and \textbf{AI systems make predictions} entry points, the educators came up with an idea about an activity around an unfair machine, where the output from the machine would favor specific users. They explained that with such a machine, children could reflect on how the data used by the machine was created and how data influenced the suggestions made by the system. The educators developed this idea into an activity where lower elementary school children would pick a game from a physical box in the classroom, which only contained games that were mainly played when their grandparents were children. In the activity the children then had to redesign the game into something they would like to play. The educators brought this idea to a workshop with researchers; there, different stakeholders discussed the potential of bridging this idea with the episode from the media company described in Vignette 2, as these both involved recommendation systems.  

In this meeting the educators and researchers brainstormed that a food recommendation system could potentially be used as an activity that could bridge the school and home contexts. Researchers then built a web app that showed how recommendation systems make predictions based on existing data. The app was designed to recommend food and games using an initial dataset skewed towards adults' ideas of what good games and food are. It was designed so that children and their parents could add new food and game ideas to the dataset and delete data points. They could also add the same idea several times to the dataset to increase the probability of getting a certain food or game recommended. 

The analogue activity and the web app were designed to bridge what children learn at school and at home. The plan is that children will first use the analogue version of the activity at school, then transition to the web app to modify the data to their own interests. Then they can bring it home and show what they have made to their parents and potentially trick them with a recommendation system they have designed with their own data and intentions. This vignette illustrates how diverse stakeholders used the entry points to design activities that complement each other and that can bridge what happens at school and at home.

\section{Discussion}
In this paper, we asked: \textbf{How can multiple stakeholders negotiate and translate AI literacy framework ideas into common language that can be used to design introductory learning materials for primary school children, teachers, and parents?} Our experience suggests that one way of addressing this challenge is through the development of entry points. We reported on how the conceptually grounded entry points became generative starting points developed through a participatory process that translated ideas from existing AI literacy and informatics frameworks into shared reference points for the design of educational materials.

Despite the fact that there are numerous frameworks to guide AI literacy efforts, educators and designers often struggle to translate these frameworks into practice \cite{yang2026systematic}. The five entry points suggested in this project provided a shared vocabulary that teacher educators, media producers, curricular designers, and researchers could use in their design processes. Rather than prescribing learning objectives, the entry points enabled stakeholders to develop different manifestations of AI literacy ideas across learning tools, educational activities, and media content. Our three vignettes illustrate this generativity: the same entry points informed the design of an explainable chatbot, narratives for children's television, and recommendation-system activities connecting school and home.

\subsection{Translating Frameworks}

The need for entry points arises from a distinction between specifying what matters for AI literacy and supporting educational designers in deciding where and how to begin. Existing frameworks provide rich accounts of competencies, concepts, and areas of knowledge. Yet, as discussed earlier, the frameworks themselves do not resolve the problem of applying them. Frameworks may remain difficult to translate into concrete educational practice because of knowledge gaps between framework authors, designers, and teacher trainers, insufficient guidance for local contexts, and the breadth and complexity of the concepts involved.

Entry points address the translation of AI frameworks into the process of designing new educational materials. They do not replace the AI frameworks, reduce them to a new taxonomy, nor provide another comprehensive account of what children should learn about AI. Instead, they offer a layer of translation between frameworks and actual design practices. We made this distinction visible in Table \ref{tab:entry}, where each entry point can be traced to \citeauthor{touretzky2019envisioning}'s \citeyearpar{touretzky2019envisioning} big ideas, \citeauthor{10.1145/3313831.3376727}'s \citeyearpar{10.1145/3313831.3376727} framework and the EU's Informatics Reference Framework \cite{10.1145/3583088}. Entry points therefore represent a deliberate reduction in comprehensiveness in exchange for generativity and actionability. They make selected ideas available as starting points around which educational materials can be imagined, discussed, designed and ultimately implemented.

This translational function was particularly important from our outset of computational empowerment (CE). As argued earlier, computational empowerment shifts attention from learning to use AI towards supporting children to understand, question, construct, deconstruct, and ultimately participate in shaping contemporary technologies. 

The entry points preserve this CE orientation in a form that can inform design. Data is created directs attention towards human practices involved in producing representations of the world; AI systems make predictions foregrounds the probabilistic rather than deterministic character of AI outputs; anthropomorphism in AI systems opens questions about how interfaces shape human understandings of machine agency; AI as a design material positions AI as something that humans can construct and modify; and intentionality in AI systems foregrounds the values, choices, and purposes embedded through design.

\subsection{A Participatory Approach}
The entry points cannot be separated from the process through which they were produced. These  were not derived by researchers from existing frameworks and subsequently handed to practitioners for implementation. Although researchers initially compared existing AI frameworks and recommendations and identified provisional topic areas, these topics were explicitly introduced into the participatory design process as starting points rather than fixed specifications. Teacher trainers, media producers, and learning-technology designers subsequently interpreted them from within their respective professional practices, and their questions and concerns shaped how the entry points were articulated.

In respect to this process, the entry points should consequently be understood as both products of participation and infrastructures for continued participation. This distinction matters. From a computational empowerment perspective, participation was not conceived as consulting practitioners about a framework already determined by researchers. Consistent with the Scandinavian PD tradition \cite{958052.958061, bodker2022participatory} underpinning computational empowerment, stakeholders need sufficient ownership to question researchers' assumptions about AI literacy and determine how framework concepts should inform their professional work. 

\subsection{Limitations}

Our paper has several limitations. First, the contribution of the entry points concerns the design process, not the effectiveness of the resulting materials. The three vignettes demonstrate how teacher educators, media creators and researchers used the entry points to create educational materials. They do not demonstrate that the entry points themselves support children's AI literacy, nor do they establish that teachers or parents will interpret the resulting materials as intended. We cannot claim that these five entry points are sufficient to foster AI literacy or CE among learners, a limitation consistent with the scope we established when introducing them. Future research should follow the materials to investigate how teachers, children, and parents appropriate them.

A second limitation concerns the cultural and institutional conditions under which the entry points emerged. CE itself is rooted in the Scandinavian participatory design tradition, and the PD process reported here similarly assumes that professional stakeholders should have meaningful opportunities to question concepts, negotiate priorities, and shape the resulting educational agenda. 
These conditions are not culturally or institutionally neutral. Rather than treating this as a problem to be resolved through prescriptive standards, it raises an important question for future research: Which aspects of entry points can travel across contexts, and which need to be renegotiated locally? 

Finally, we cannot claim that the five entry points are exhaustive or universally applicable. Rather, they should be understood as situated outcomes of the particular participatory process through which they emerged. Their formulation reflects the specific stakeholders, practices, and concerns brought together in this project. In our case computational empowerment heavily influenced the development of the entry points. Other PD processes, involving different stakeholders or situated in different contexts, may lead to the development of different entry points. 

\subsection{Contribution}
Despite these limitations, our contribution extends beyond the five specific entry points presented in this paper. We reported on how PD can provide a process for translating comprehensive AI literacy frameworks into generative and negotiable resources for designing educational materials. For researchers, entry points offer a way of bridging frameworks and situated design practice without prescribing how frameworks should be put into action. Thus, rather than adopting our five entry points as a fixed set, we invite researchers to consider how and what entry points they might collaboratively develop together with their partners.

\section*{Acknowledgments}
We are grateful to our partners at the Danish Broadcasting Corporation (DR), Center for Undervisningsmidler (CFU), and LIFE Fonden. This work was possible through the generous support of the Novo Nordisk Foundation and the Villum Foundation.

\bibliography{aaai2027}

@techreport{OECD2026EmpoweringLearners,
  author      = {{OECD} and {European Union}},
  title       = {Empowering learners for the age of AI: An AI literacy framework for primary and secondary education},
  year        = {2026},
  institution = {OECD Publishing},
  address     = {Paris},
  doi         = {10.1787/65cd27d4-en},
  url         = {https://doi.org/10.1787/65cd27d4-en}
}

@book{unesco2024aicompetency,
  title     = {AI Competency Framework for Students},
  author    = {Miao, Fengchun and Shiohira, Kelly and Lao, Natalie},
  organization = {UNESCO},
  year      = {2024},
  publisher = {UNESCO},
  isbn      = {978-92-3-100709-5},
  doi       = {10.54675/JKJB9835},
  url       = {https://doi.org/10.54675/JKJB9835},
  pages     = {80},
  language  = {English},
  note      = {Licence: CC BY-SA 3.0 IGO}
}

@article{yeter2024global,
  title={Global initiatives and challenges in integrating artificial intelligence literacy in elementary education: Mapping policies and empirical literature},
  author={Yeter, Ibrahim H and Yang, Weipeng and Sturgess, Joshita B},
  journal={Future in Educational Research},
  volume={2},
  number={4},
  pages={382--402},
  year={2024},
  publisher={Wiley Online Library}
}

@incollection{star1989structure,
  title={The structure of ill-structured solutions: Boundary objects and heterogeneous distributed problem solving},
  author={Star, Susan Leigh},
  booktitle={Distributed artificial intelligence},
  pages={37--54},
  year={1989},
  publisher={Elsevier}
}

@article{10.1145/3583088,
author = {Caspersen, Michael and Gal-Ezer, Judith and McGettrick, Andrew and Nardelli, Enrico},
title = {Informatics Education for School: A European Initiative},
year = {2023},
issue_date = {March 2023},
publisher = {Association for Computing Machinery},
address = {New York, NY, USA},
volume = {14},
number = {1},
issn = {2153-2184},
url = {https://doi.org/10.1145/3583088},
doi = {10.1145/3583088},
journal = {ACM Inroads},
month = feb,
pages = {49–53},
numpages = {5}
}

@article{958052.958061,
author = {B{\o}dker, Susanne},
title = {A for alternatives},
year = {2003},
issue_date = {01/01/2003},
publisher = {University of Aalborg},
address = {DNK},
volume = {15},
number = {1},
issn = {0905-0167},
journal = {Scand. J. Inf. Syst.},
month = jan,
pages = {87–89},
numpages = {3}
}

@article{casal2023ai,
  title={AI literacy in K-12: a systematic literature review},
  author={Casal-Otero, Lorena and Catala, Alejandro and Fern{\'a}ndez-Morante, Carmen and Taboada, Maria and Cebreiro, Beatriz and Barro, Sen{\'e}n},
  journal={International journal of STEM education},
  volume={10},
  number={1},
  pages={29},
  year={2023},
  publisher={Springer}
}

@article{yang2026systematic,
  title={A systematic review mapping of AI literacy progression in K--12},
  author={Yang, Hui and Rachmatullah, Arif and Alozie, Nonye and Capan, Selin and Cao, Qican},
  journal={Journal for STEM Education Research},
  pages={1--32},
  year={2026},
  publisher={Springer}
}

@article{dindler2022computational,
  title={Computational empowerment},
  author={Dindler, Christian and Iversen, Ole Sejer and Caspersen, Michael E and Smith, Rachel Charlotte},
  journal={Computational thinking education in K--12: Artificial intelligence literacy and physical computing},
  pages={121--139},
  year={2022},
  publisher={The MIT Press Cambridge}
}

@techreport{InformaticsForAll2022Framework,
  author      = {Caspersen, Michael E. and Diethelm, Ira and Gal-Ezer, Judith and McGettrick, Andrew and Nardelli, Enrico and Passey, Don and Rovan, Branislav and Webb, Mary},
  title       = {The Informatics Reference Framework for School},
  institution = {Informatics for All Coalition},
  month       = feb,
  year        = {2022},
  url         = {https://www.informaticsforall.org/the-informatics-reference-framework-for-school-online-en/}
}

@book{bodker2022participatory,
  author    = {B{\o}dker, Susanne and Dindler, Christian and Iversen, Ole S. and Smith, Rachel C.},
  title     = {Participatory Design},
  series    = {Synthesis Lectures on Human-Centered Informatics},
  publisher = {Springer Cham},
  year      = {2022},
  edition   = {1},
  isbn      = {978-3-031-01107-8},
  eisbn     = {978-3-031-02235-7},
  issn      = {1946-7680},
  eissn     = {1946-7699},
  doi       = {10.1007/978-3-031-02235-7}
}

@article{mutawa2025unesco,
  title={UNESCO's AI competency framework: Challenges and opportunities in educational settings},
  author={Mutawa, AM and Sruthi, Sai},
  journal={Impacts of Generative AI on the Future of Research and Education},
  pages={75--96},
  year={2025},
  publisher={IGI Global Scientific Publishing}
}

@inproceedings{touretzky2019envisioning,
  title={Envisioning {AI} for {K-12}: What should every child know about AI?},
  author={Touretzky, David and Gardner-McCune, Christina and Martin, Fred and Seehorn, Deborah},
  booktitle={Proceedings of the AA{AI} conference on artificial intelligence},
  volume={33},
  pages={9795--9799},
  year={2019}
}

@inproceedings{xu2026essentials,
  title={The essentials of AI for life and society: A full-scale AI literacy course accessible to all},
  author={Xu, Zifan and Procko, Kristen and Munje, Michael and Patterson, Kristin and Sabatini, Lea and Biswas, Joydeep and Stone, Peter},
  booktitle={Proceedings of the AAAI Conference on Artificial Intelligence},
  volume={40},
  number={47},
  pages={40730--40736},
  year={2026}
}

@inproceedings{10.1145/3769994.3770026,
author = {Vartiainen, Henriikka and Tedre, Matti},
title = {The CEDE Model: A Learning-Sciences Based Approach for Critical and Transformative K–12 AI Education},
year = {2025},
isbn = {9798400715990},
publisher = {Association for Computing Machinery},
address = {New York, NY, USA},
url = {https://doi.org/10.1145/3769994.3770026},
doi = {10.1145/3769994.3770026},
booktitle = {Proceedings of the 25th Koli Calling International Conference on Computing Education Research},
articleno = {36},
numpages = {11},
location = {
},
series = {Koli Calling '25}
}

@inproceedings{10.1145/3773077.3806126,
author = {Madushika Jayathilake, Hansinie and Ma, Renkai},
title = {Anthropomorphism in Children's Interactions with LLM-Based Chatbots: A Systematic Review of Drivers and Outcomes},
year = {2026},
isbn = {9798400722837},
publisher = {Association for Computing Machinery},
address = {New York, NY, USA},
url = {https://doi.org/10.1145/3773077.3806126},
doi = {10.1145/3773077.3806126},
booktitle = {Proceedings of the 25th Annual ACM Interaction Design and Children Conference},
pages = {520–536},
numpages = {17},
location = {
},
series = {IDC '26}
}

@inproceedings{10.1145/3702652.3744217,
author = {Gu, Xingjian (Lance) and Ericson, Barbara J.},
title = {{AI} Literacy in {K-12} and Higher Education in the Wake of Generative AI: An Integrative Review},
year = {2025},
isbn = {9798400713408},
publisher = {Association for Computing Machinery},
address = {New York, NY, USA},
doi = {10.1145/3702652.3744217},
booktitle = {Proceedings of the 2025 ACM Conference on International Computing Education Research V.1},
pages = {125–140},
numpages = {16},
location = {
},
series = {ICER '25}
}

@inproceedings{10.1145/3210586.3210592,
author = {Iversen, Ole Sejer and Smith, Rachel Charlotte and Dindler, Christian},
title = {From computational thinking to computational empowerment: a 21st century PD agenda},
year = {2018},
isbn = {9781450363716},
publisher = {Association for Computing Machinery},
address = {New York, NY, USA},
url = {https://doi.org/10.1145/3210586.3210592},
doi = {10.1145/3210586.3210592},
booktitle = {Proceedings of the 15th Participatory Design Conference: Full Papers - Volume 1},
articleno = {7},
numpages = {11},
location = {Hasselt and Genk, Belgium},
series = {PDC '18}
}

@article{waite2023ethics,
  title={From ethics to engines, framing AI literacy (including algorithm literacy and data literacy) through the SEAME framework},
  author={Waite, Jane and Garside, Ben},
  journal={Raspberry Pi Computing Education Research Centre, Cambridge United Kingdom},
  year={2023}
}

@article{kalantzis2025literacy,
  title={Literacy in the time of Artificial Intelligence},
  author={Kalantzis, Mary and Cope, Bill},
  journal={Reading Research Quarterly},
  volume={60},
  number={1},
  pages={e591},
  year={2025},
  publisher={Wiley Online Library}
}

@inproceedings{iversen2017child,
  author = {Iversen, Ole Sejer and Smith, Rachel Charlotte and Dindler, Christian},
title = {Child as Protagonist: Expanding the Role of Children in Participatory Design},
year = {2017},
isbn = {9781450349215},
publisher = {Association for Computing Machinery},
address = {New York, NY, USA},
url = {https://dl.acm.org/doi/10.1145/3078072.3079725},
doi = {10.1145/3078072.3079725},
booktitle = {Proceedings of the 2017 Conference on Interaction Design and Children},
pages = {27–37},
numpages = {11},
location = {Stanford, California, USA},
series = {IDC '17}
}

@inproceedings{10.1145/3772318.3791296,
author = {Liu, Yu-Yu and Dybboe, Maja and Ellemose, Johannes and Bjerrum, Johanne Birkkj{\ae}r and Musaeus, Line Have and Dindler, Christian and Iversen, Ole Sejer},
title = {"Let’s talk about data": Co-Designing Critical Data Literacy Tools for K-12 Education through Dialogic Learning},
year = {2026},
isbn = {9798400722783},
publisher = {Association for Computing Machinery},
address = {New York, NY, USA},
url = {https://doi.org/10.1145/3772318.3791296},
doi = {10.1145/3772318.3791296},
booktitle = {Proceedings of the 2026 CHI Conference on Human Factors in Computing Systems},
articleno = {92},
numpages = {23},
location = {
},
series = {CHI '26}
}

@article{10.1145/3757620,
author = {Morales-Navarro, Luis and Gan, Michelle A. and Yu, Evelyn and Vogelstein, Lauren and Kafai, Yasmin and Metaxa, Dana{\'e}},
title = {Learning AI Auditing: A Case Study of Teenagers Auditing a Generative AI Model},
year = {2025},
issue_date = {November 2025},
publisher = {Association for Computing Machinery},
address = {New York, NY, USA},
volume = {9},
number = {7},
url = {https://doi.org/10.1145/3757620},
doi = {10.1145/3757620},
journal = {Proc. ACM Hum.-Comput. Interact.},
month = oct,
articleno = {CSCW439},
numpages = {29}
}

@incollection{kahila2026engaging,
  title={Engaging young minds in AI through co-designing and making their own AI-Driven Apps in the classroom},
  author={Kahila, Juho and Vartiainen, Henriikka and Pope, Nicolas and Tedre, Matti},
  booktitle={Social robots and artificial intelligence in education: Integrating AI in K-12 and higher education},
  pages={187--210},
  year={2026},
  publisher={Springer}
}

@incollection{erickson2012qualitative,
  author    = {Erickson, Frederick},
  title     = {Qualitative Research Methods for Science Education},
  booktitle = {Second International Handbook of Science Education},
  editor    = {Fraser, Barry and Tobin, Kenneth and McRobbie, Campbell},
  series    = {Springer International Handbooks of Education},
  volume    = {24},
  publisher = {Springer},
  address   = {Dordrecht},
  year      = {2012},
  pages     = {1451--1465},
  doi       = {10.1007/978-1-4020-9041-7_93},
  url       = {https://doi.org/10.1007/978-1-4020-9041-7_93}
}

@inproceedings{hilke2026breakable,
  title={Breakable Machine: AK--12 classroom game for transformative AI literacy through spoofing and explainable AI (XAI)},
  author={Hilke, Olli and Pope, Nicolas and Kahila, Juho and Vartiainen, Henriikka and Roos, Teemu and Parkki, Tuomo and Tedre, Matti},
  booktitle={Proceedings of the AAAI Conference on Artificial Intelligence},
  volume={40},
  number={47},
  pages={40762--40770},
  year={2026}
}

@INPROCEEDINGS{11574161,
  author={Moilanen, Markus and Hilke, Olli and Kahila, Juho and Parkki, Tuomo and Silvennoinen, Elisa and Pope, Nicolas and Tedre, Matti},
  booktitle={2026 IEEE Global Engineering Education Conference (EDUCON)}, 
  title={Learning Machine Learning Through Deconstructing (X)AI Systems in K–12 Education}, 
  year={2026},
  volume={},
  number={},
  pages={1-5},
  doi={10.1109/EDUCON67543.2026.11574161}}

@inproceedings{10.1145/3531146.3534634,
author = {Bilstrup, Karl-Emi Kj{\ae}r and Kaspersen, Magnus H{\o}holt and Assent, Ira and Enni, Simon and Petersen, Marianne Graves},
title = {From Demo to Design in Teaching Machine Learning},
year = {2022},
isbn = {9781450393522},
publisher = {Association for Computing Machinery},
address = {New York, NY, USA},
url = {https://doi.org/10.1145/3531146.3534634},
doi = {10.1145/3531146.3534634},
booktitle = {Proceedings of the 2022 ACM Conference on Fairness, Accountability, and Transparency},
pages = {2168–2178},
numpages = {11},
location = {Seoul, Republic of Korea},
series = {FAccT '22}
}

@inproceedings{10.1145/3357236.3395558,
author = {Bilstrup, Karl-Emil Kj{\ae}r and Kaspersen, Magnus H. and Petersen, Marianne Graves},
title = {Staging Reflections on Ethical Dilemmas in Machine Learning: A Card-Based Design Workshop for High School Students},
year = {2020},
isbn = {9781450369749},
publisher = {Association for Computing Machinery},
address = {New York, NY, USA},
url = {https://doi.org/10.1145/3357236.3395558},
doi = {10.1145/3357236.3395558},
booktitle = {Proceedings of the 2020 ACM Designing Interactive Systems Conference},
pages = {1211–1222},
numpages = {12},
location = {Eindhoven, Netherlands},
series = {DIS '20}
}

@inproceedings{10.1145/3773077.3806141,
author = {Connelly, Luke and Graves Petersen, Marianne and H{\o}holt Kaspersen, Magnus and Musaeus, Line Have and Bilstrup, Karl-Emil Kj{\ae}r},
title = {Addressing the NLP in the Classroom: Empowering High School Teachers through Participatory Design of Hands-on NLP Activities},
year = {2026},
isbn = {9798400722837},
publisher = {Association for Computing Machinery},
address = {New York, NY, USA},
url = {https://doi.org/10.1145/3773077.3806141},
doi = {10.1145/3773077.3806141},
booktitle = {Proceedings of the 25th Annual ACM Interaction Design and Children Conference},
pages = {688–702},
numpages = {15},
location = {
},
series = {IDC '26}
}

@inproceedings{10.1145/3773077.3806107,
author = {Morales-Navarro, Luis and Noh, Daniel J. and Servat, Lucianne and Netting, Carly and Kafai, Yasmin and Metaxa, Dana{\'e}},
title = {Building to Understand: Examining Teens’ Technical and Socio-Ethical Pieces of Understanding in the Construction of Small Generative Language Models},
year = {2026},
isbn = {9798400722837},
publisher = {Association for Computing Machinery},
address = {New York, NY, USA},
url = {https://doi.org/10.1145/3773077.3806107},
doi = {10.1145/3773077.3806107},
booktitle = {Proceedings of the 25th Annual ACM Interaction Design and Children Conference},
pages = {262–280},
numpages = {19},
location = {
},
series = {IDC '26}
}

@inproceedings{10.1145/3459990.3460712,
author = {Druga, Stefania and Ko, Amy J},
title = {How do children’s perceptions of machine intelligence change when training and coding smart programs?},
year = {2021},
isbn = {9781450384520},
publisher = {Association for Computing Machinery},
address = {New York, NY, USA},
url = {https://doi.org/10.1145/3459990.3460712},
doi = {10.1145/3459990.3460712},
booktitle = {Proceedings of the 20th Annual ACM Interaction Design and Children Conference},
pages = {49–61},
numpages = {13},
location = {Athens, Greece},
series = {IDC '21}
}

@inproceedings{10.1145/3547522.3547689,
author = {Bilstrup, Karl-Emil Kj{\ae}r and Kaspersen, Magnus H{\o}holt and S{\o}rensen, Marie-Louise Stisen Kjerstein and Petersen, Marianne Graves},
title = {Opportunities and Challenges of Teaching Machine Learning as a Design Material with the micro:bit},
year = {2022},
isbn = {9781450394482},
publisher = {Association for Computing Machinery},
address = {New York, NY, USA},
url = {https://doi.org/10.1145/3547522.3547689},
doi = {10.1145/3547522.3547689},
booktitle = {Adjunct Proceedings of the 2022 Nordic Human-Computer Interaction Conference},
articleno = {45},
numpages = {6},
location = {Aarhus, Denmark},
series = {NordiCHI '22 Adjunct}
}

@article{hardy2020data,
  title={From data collectors to data producers: Shifting students’ relationship to data},
  author={Hardy, Lisa and Dixon, Colin and Hsi, Sherry},
  journal={Journal of the Learning Sciences},
  volume={29},
  number={1},
  pages={104--126},
  year={2020},
  publisher={Taylor \& Francis}
}

@inproceedings{10.1145/3313831.3376727,
author = {Long, Duri and Magerko, Brian},
title = {What is AI Literacy? Competencies and Design Considerations},
year = {2020},
isbn = {9781450367080},
publisher = {Association for Computing Machinery},
address = {New York, NY, USA},
url = {https://doi.org/10.1145/3313831.3376727},
doi = {10.1145/3313831.3376727},
booktitle = {Proceedings of the 2020 CHI Conference on Human Factors in Computing Systems},
pages = {1–16},
numpages = {16},
location = {Honolulu, HI, USA},
series = {CHI '20}
}

@inproceedings{10.1145/3626252.3630937,
author = {Grover, Shuchi},
title = {Teaching {AI} to {K-12} Learners: Lessons, Issues, and Guidance},
year = {2024},
isbn = {9798400704239},
publisher = {Association for Computing Machinery},
address = {New York, NY, USA},
url = {https://doi.org/10.1145/3626252.3630937},
doi = {10.1145/3626252.3630937},
booktitle = {Proceedings of the 55th ACM Technical Symposium on Computer Science Education V. 1},
pages = {422–428},
numpages = {7},
location = {Portland, OR, USA},
series = {SIGCSE 2024}
}

@inproceedings{kasinidou2026educating,
  title={Educating the Public in Artificial Intelligence: Insights from a Large-Scale Course},
  author={Kasinidou, Maria and Kleanthous, Styliani and Otterbacher, Jahna and Christoforou, Evgenia},
  booktitle={Proceedings of the AAAI Conference on Artificial Intelligence},
  volume={40},
  number={47},
  pages={40617--40625},
  year={2026}
}

@inproceedings{Whyte2024_secondary_AI_conception,
author = {Whyte, Robert and Kirby, Diana and Sentance, Sue},
title = {Secondary Students' Emerging Conceptions of AI: Understanding AI Applications, Models, Engines and Implications},
year = {2024},
isbn = {9798400711770},
publisher = {Association for Computing Machinery},
address = {New York, NY, USA},
url = {https://doi.org/10.1145/3689535.3689552},
doi = {10.1145/3689535.3689552},
booktitle = {Proceedings of the 2024 Conference on United Kingdom \& Ireland Computing Education Research},
articleno = {3},
numpages = {7},
location = {Manchester, United Kingdom},
series = {UKICER '24}
}

@article{cheng2026metaphors,
  title={Metaphors of AI indicate that people increasingly perceive AI as warm and human-like},
  author={Cheng, Myra and Lee, Angela Y and Rapuano, Kristina and Niederhoffer, Kate and Liebscher, Alex and Hancock, Jeffrey},
  journal={Communications Psychology},
  volume={4},
  number={1},
  pages={8},
  year={2026},
  publisher={Nature Publishing Group UK London}
}


\end{document}